\documentclass[aps,prmaterials,preprint,floatfix]{revtex4-1}
\usepackage{amsmath}
\usepackage{amsfonts}
\usepackage{amssymb}
\usepackage{bm}
\def\*#1{\mathbf{#1}}
\usepackage[usenames,dvipsnames]{color} 
\usepackage{ulem}
\usepackage{graphicx}
\graphicspath{ {./figs/} }
\usepackage{epstopdf}

\usepackage{xcolor}

\usepackage{chemformula}
\usepackage{csquotes}
\usepackage[sort&compress]{natbib}
\usepackage [english]{babel}
\usepackage[utf8]{inputenc}
\usepackage[T1]{fontenc}

\usepackage{gensymb}

\usepackage{lineno}

\usepackage{hyperref}
\hypersetup{
  colorlinks   = true, 
  urlcolor     = black, 
  linkcolor    = red, 
  citecolor   = blue 
}

\usepackage{tabularx}
\newcolumntype{Y}{>{\centering\arraybackslash}X}  

\begin{document}

\title{Fracture of silicate glasses: Micro-cavities and correlations between atomic-level properties
}

\author{Zhen Zhang}

\affiliation{Laboratoire Charles Coulomb (L2C), 
University of Montpellier and CNRS, F-34095 Montpellier, France}

\author{Simona Ispas}
\affiliation{Laboratoire Charles Coulomb (L2C), 
University of Montpellier and CNRS, F-34095 Montpellier, France}

\author{Walter Kob}
\email[Corresponding author: ]{walter.kob@umontpellier.fr}
\affiliation{Laboratoire Charles Coulomb (L2C),
University of Montpellier and CNRS, F-34095 Montpellier, France}

\date{\today}

\begin{abstract}  

We use large-scale simulations to investigate the dynamic fracture of silica and sodium-silicate glasses under uniaxial tension. The stress-strain curves demonstrate that silica glass is brittle whereas the glasses rich in Na show pronounced ductility. A strong composition dependence is also seen in the crack velocity which is on the order of 1800~m/s for glasses with low Na concentration and decreases to 700~m/s if the concentration is high. We find that during the fracture of Na-rich glasses very irregular cavities as large as 3-4 nm form ahead of the crack front, indicating the presence of nanoductility in these glasses.  Before fracture occurs, the local composition, structure, and mechanical properties are heterogeneous in space and show a strong dependence on the applied strain. Further analysis of the correlations between these local properties allows to obtain a better microscopic understanding of the deformation and fracture of glasses and how the local heating close to the crack tip, up to several hundred degrees, 
permits the structure to relax.


\end{abstract}


\maketitle
\section{Introduction}
Silicate glasses constitute a class of amorphous materials 
that is not only of high scientific interest but is also most important for many technical applications~\cite{binder_glassy_2011,shelby_introduction_2005,
varshneya2013fundamentals}. For many of these applications, the mechanical reliability of the material is crucial and hence this subject has been in the focus of interest of numerous studies in glass science and technology~\cite{wondraczek_towards_2011,wondraczek2022review}. Silicate glasses are known to be archetypal brittle materials which show little or no plastic deformation before fracture, although they do display a pronounced non-linear elastic behavior~\cite{mallinder1964elastic,krause_deviations_1979,gupta_intrinsic_2005,guerette_nonlinear_2016,zhang2022stiffness}. However, whether or not this well-known property is also observed  on the micro-and meso-scale has been controversial for decades, as documented in several review articles~\cite{tomozawa1996fracture,ciccotti_stress-corrosion_2009,wiederhorn_griffith_2013}. One of the questions that is highly debated is whether or not the fracture of oxide glasses is accompanied by the formation, growth and coalescence of microscopic cavities in front of the crack tip~\cite{rountree2002atomistic,celarie_glass_2003,guin_fracture_2004,bonamy_nanoscale_2006,muralidharan_molecular_2005,rountree_unified_2007,pedone_molecular_2008,pedone_dynamics_2015,wang_intrinsic_2015,shen2021observation}, features that are usually associated with ductile fracture. In experiments glass fracture is frequently investigated by means of atomic force microscopy (AFM), often in a stress-corrosive environment, i.e., when water interacts with the glass structure~\cite{wiederhorn_use_2011}. This experimental setup allows to control the velocity for the crack propagation over a wide range, but usually smaller than 10$^{-2}$ m/s (thus sub-critical fracture), and permits the direct observation of crack propagation \textit{in situ}~\cite{ciccotti_situ_2018}. Cavities with sizes ranging from 5~nm to more than 100~nm have been reported, and this size depends on various factors, such as the measuring method, stress condition, and crack velocity, although the exact dependence on these factors is not really understood~\cite{celarie_glass_2003,guin_fracture_2004,bonamy_nanoscale_2006,waurischk2021vacuum}. Also molecular dynamics (MD) simulations have been used to study the fracture of silicate glasses, usually in a vacuum environment. These studies reported the existence of voids having sizes around 0.5 nm, weakly depending on glass composition~\cite{wang_intrinsic_2015,wang_nanoductility_2016,
pedone_molecular_2008,pedone_dynamics_2015}. However, the comparison between such simulations and the experimental studies of stress-corrosive fracture has to be done with caution, since these two techniques differ strongly in the relevant parameters such as crack velocities, chemical conditions, strain rates, etc. Hence at the end our microscopic understanding of the fracture dynamics in oxide glasses is not at all in a satisfying state.

A further important question concerns the modification of the local properties of the glass during fracture. Since close to the crack tip the stresses are high, one can expect that in this region the structure and the mechanical properties are very different from the ones in the bulk. Unfortunately experimental techniques do not allow to probe these quantities with high precision since the 
process zone near the crack tip in which energy dissipation occurs has been reported to have a size of less than 10~nm~\cite{han_measuring_2010,pallares2012fractoluminescence}, i.e., is at the resolution limit of typical experimental probes. Recent atomistic simulations have allowed to gain some insight into  atomic-scale properties, such as the dissipation of energy in the vicinity of the crack tip using the atomistic J-integral approach~\cite{rimsza_crack_2018,rimsza2018chemical,chowdhury_effects_2019}, as well as the heterogeneities in local structure and mechanical properties and the correlations between them~\cite{wang_intrinsic_2015,wang_nanoductility_2016,hao_atomistic_2019,frankberg2019highly,du2021predicting}.  
However, at present there is still a lack of understanding on how various local properties evolve and correlate with each other as a function of strain and hence how these correlations influence the fracture behavior of oxide glasses. 
 
The goal of the present study is to obtain a microscopic understanding of the fracture behavior of silicate glasses at conditions encountered in real life, i.e., glass breaks when a surface flaw meets tension. This situation corresponds to dynamic fracture, i.e., the speed of the crack front is so fast that environment factors, such as humidity, do not influence the properties of the crack tip and hence the fracture dynamics. For this case it should therefore be realistic to model the fracture process of the sample in vacuum. We study this dynamic fracture behavior by means of large-scale MD simulations of sodium silicate glasses with different Na content. In particular we focus on the possible presence of nanoscale cavities during fracture, a question which has been debated intensively in the past but for which so far no conclusive answer has been obtained. Furthermore we determine, for the case of a modifier-rich glass, the evolution of various atomic-scale properties focusing to unravel the relationships between local composition, structure, and mechanical properties, hence obtaining new insight on the fracture mechanisms of silicate glasses on the microscopic scales.

The rest of the paper is organized as follows. In Sec. \ref{method} we introduce the simulation setups. In Sec.~\ref{results} we present and discuss the results regarding the cavitation process and various atomic-scale properties during deformation. Finally, we summarize and conclude this work in Sec.~\ref{summary}. 

\section{Methodology} \label{method}
We performed molecular dynamics simulations for a series of sodium silicates samples with the composition SiO$_2$ and Na$_2$O-$x$SiO$_2$ (denoted in the following by
NS$x$), with $x=3, 4, 5, 7, 10,$ and 20, corresponding to the variation of Na$_2$O from 0 to 25 mole \%. The interactions between the atoms were given by the pairwise effective potential proposed by Sundararaman 
et al.~\cite{sundararaman_new_2018,sundararaman_new_2019}. The functional form of this potential consists of a Buckingham term for short-range interactions and a Coulombic term which is evaluated by the Wolf truncation method~\cite{wolf_exact_1999}. More details regarding the development of the potential and its ability to reproduce experimental data and \textit{ab initio} calculations are given in Refs.~\cite{sundararaman_new_2018,sundararaman_new_2019}. Recent studies have shown that this potential allows for a good quantitative description
of the structure, mechanical, as well as surface properties of sodium silicate
glasses~\cite{zhang_potential_2020,zhang_surf-vib_2020,zhang2021roughness}, and therefore it can be expected that it is also able to give a realistic description of the fracture process.

For each sample, approximately $2,300,000$ atoms were placed randomly in the simulation box which had a fixed
volume determined by the experimental value of glass density at
room temperature~\cite{bansal_handbook_1986}, corresponding to boxes with typical dimensions  of 20~nm$\times$30~nm$\times$50~nm. Using periodic boundary
conditions in three dimensions, these samples were first melted and
equilibrated at 6000~K for 80~ps in the canonical ensemble ($NVT$)
and then cooled and equilibrated at a lower temperature $T_1$ (still
in the liquid state) for another 160~ps.  The temperature $T_1$ ranges
from 3000~K for SiO$_2$ to 2000~K for NS3 (25 mole~\% Na$_2$O), see
Ref.~\cite{zhang_surf-vib_2020,zhang_thesis_2020} for details.  Subsequently we cut the sample orthogonal to the $z-$axis, and inserted an empty space, thus creating two free surfaces, i.e.,~the sample had the geometry of a slab. Periodic boundary conditions were applied in all three directions. In order to
ensure that the two free surfaces do not interact with each other, the
thickness of the vacuum layer varied from 6~nm for silica to 14~nm for
NS3. These samples were then equilibrated at $T_1$ for 1.6~ns, a time
span that is sufficiently long to allow the reconstruction of the surfaces
and the equilibration of the interior of the samples. Subsequently the
liquid samples were cooled via a two-stage quenching: A cooling rate
of $\gamma_1=0.125$~K/ps was used to quench the samples from $T_1$ to
a temperature $T_2$ and a faster cooling rate $\gamma_2=0.375$~K/ps
to cool them from $T_2$ to 300~K. Finally, the samples were annealed
at 300~K for 800~ps. The temperature $T_2$ at which the cooling rate
changes was chosen to be at least 200~K below the simulation glass
transition temperature $T_g$, i.e.,~depending on 
composition, 1500~K $\geq T_2 \geq 800$~K, 
see Refs.~\cite{zhang_surf-vib_2020,zhang_thesis_2020} for details.  At $T_2$, we also switched the simulation ensemble from $NVT$ to $NPT$ (at zero pressure)
so that the generated glass samples were not under macroscopic stress
at room temperature. 

After the preparation of the glass samples we introduced on one of its
free surfaces a linear ``scratch'' in the form of a triangular notch spanning
the sample in the $y$-direction of width and depth of 3~nm and 2~nm,
respectively. Subsequently we applied to the sample a strain in the $x$-direction, with a
constant rate of 0.5~ns$^{-1}$, until it broke. Due to the presence of the
notch, the place at which the fracture initiated could be changed at
will. More details can be found in Ref.~\cite{zhang_thesis_2020}.

Temperature and pressure were controlled using a Nos\'e-Hoover thermostat and
barostat~\cite{nose_unified_1984,hoover_canonical_1985,hoover_constant-pressure_1986}. All simulations were carried out using the Large-scale Atomic/Molecular Massively Parallel Simulator software (LAMMPS)~\cite{plimpton_fast_1995} with a time step of 1.6~fs.

The results presented in this study correspond to one melt-quench
sample for each composition. However, we emphasize that the system sizes
considered are sufficiently large to make sample-to-sample
fluctuations negligible. A systematic evaluation of the influence of various simulation parameters (system size, cooling rate, strain rate, etc.) on the mechanical behavior has been performed and documented elsewhere~\cite{zhang_thesis_2020}. The results presented in this paper correspond to system sizes for which the fracture behavior have basically become independent of the system size.
 
\section{Results and discussion}
\label{results}
\subsection{Stress-strain behavior}

\begin{figure}[ht]
\centering
\includegraphics[width=0.95\columnwidth]{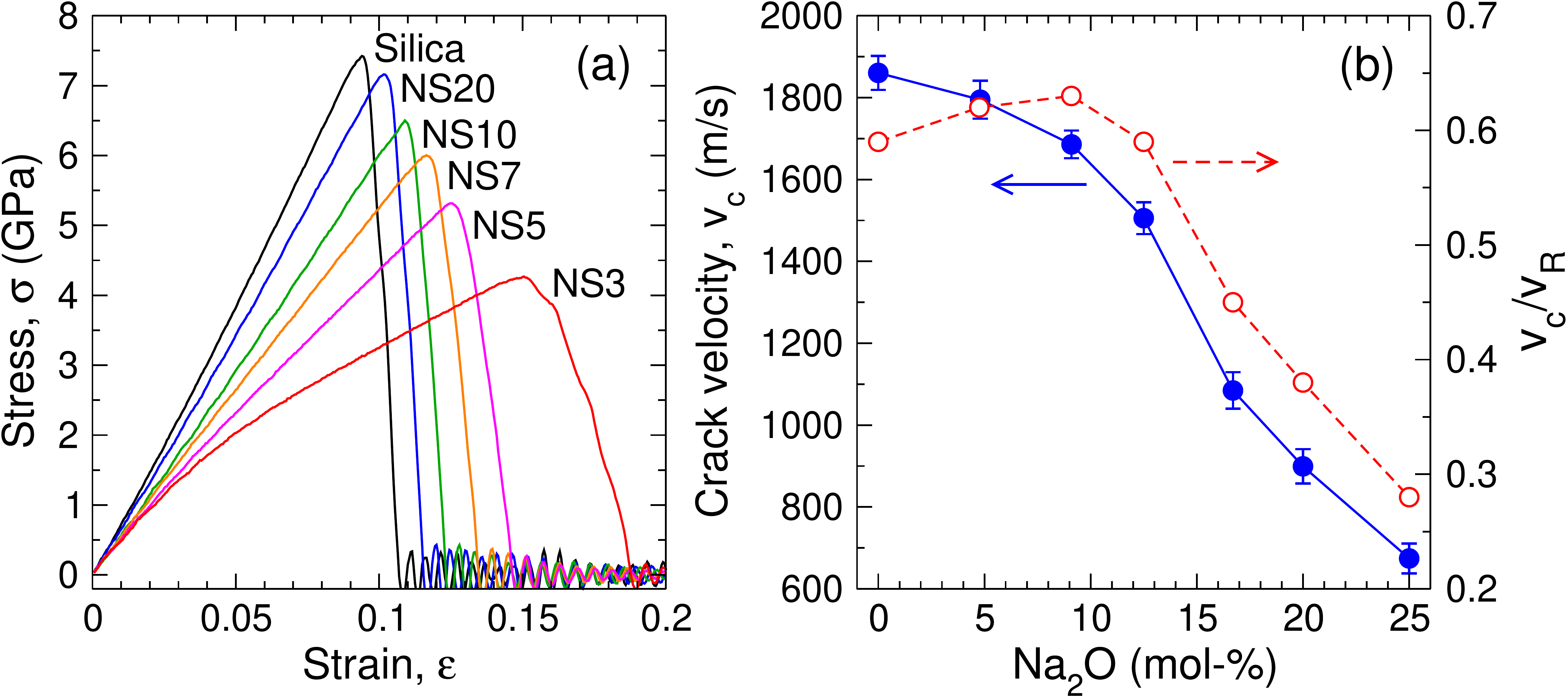}
\caption{(a) Stress-strain behavior of the NS$x$ glasses. Note that because of the periodic boundary conditions in the tensile direction, the elastic waves generated by the fracture will bounce back and forth between the two surfaces. This explains the small oscillations in the stress-strain curves after fracture. (b) Crack velocity $v_c$ (filled symbols) and the ratio between $v_c$ and the estimated Rayleigh wave speed $v_R$ (open symbols, right scale). 
}
\label{fig1_ss-crack-velo}
\end{figure}

We first demonstrate that the composition of the glass has a strong influence on the strength and fracture dynamics of the samples. In Fig.~\ref{fig1_ss-crack-velo}a we present the stress-strain curves for the different glasses and one sees a clear dependence of these curves on the concentration of sodium. These curves are qualitatively similar to the ones presented in Ref.~\cite{zhang2022stiffness} for the case of bulk systems, but we find that here the failure strain and strength (corresponding to the point of maximum stress) of the glasses are considerably smaller than the corresponding values for the bulk systems~\cite{zhang_potential_2020} or in experiments~\cite{lower_inert_2004}. This difference can be attributed to the fact that in certain experiments and simulations of the bulk systems one probes the intrinsic values of the failure strain and strength, while here the fracture is strongly affected by the presence of the surface flaw. 

The stress-strain curves demonstrate that silica glass is stiff (i.e., high elastic modulus) and brittle (i.e., abrupt stress drop after the maximum strength). The glass becomes softer and more ductile (i.e., can withstand more deformation before failure) upon increasing the concentration of Na$_2$O. Related to this behavior is the fact that the fracture process becomes more gentle with the increase of sodium concentration as can be quantified via the crack velocity $v_c$. This velocity can be estimated via the expression
$v_c = L_c/(t_0-t_m)$,
where $L_c$ denotes the total crack length and $t_0-t_m$ is the duration of crack propagation, i.e., stress drops from the maximum (at $t_m$) to zero (at time $t_0$). The dependence of $v_c$ on the NaO$_2$ concentration is shown in Fig.~\ref{fig1_ss-crack-velo}b and one sees that $v_c$ decreases from $\approx$1800~m/s for silica to $\approx$650~m/s for NS3. Note that the crack velocity depends on various factors such as composition and measurement method and hence it is difficult to make a quantitative comparison between experimental and simulation results. The crack velocities measured in our simulations are on the order of 10$^3$ m/s which is what one typically finds in experiments probing the dynamic fracture of silicate glasses~\cite{quinn2019terminal,quinn_fracture_2017,bradt_applying_2014,
pallares2012fractoluminescence}.
Our data show that for low concentration of Na this dependence is rather mild but it starts to accelerate once the concentration exceeds 10~mol~\%. A previous study has found that at around this threshold the glass samples have a significant change in their non-linear elastic properties~\cite{zhang2022stiffness} and hence it can be expected that these changes influence also the magnitude of the crack velocity. 

It is instructive to compare the values of $v_c$ with the sound velocities of silicate glasses. For pure silica at room temperature the experimental values for the transverse and longitudinal sound velocities are, respectively, 3743~m/s and 5953~m/s~\cite{polian2002elastic} and the ones for sodium silicate glasses are a few percents lower~\cite{zhao_-situ_2012,zhang_thesis_2020}. 
Hence one concludes that the crack velocities are significantly lower than the speed of sound, but of course still orders of magnitude faster than the sub-critical fracture processes studies in experiments~\cite{ciccotti_situ_2018}. Furthermore we mention that the crack velocity generally increases with strain rate, but this dependence is weak, although somewhat more pronounced for Na-rich compositions~\cite{zhang_thesis_2020}.

Within the theory of linear elastic fracture mechanics one finds that mode I (tensile) cracks cannot propagating faster than the Rayleigh wave speed, $v_R$, since this would correspond to a negative energy release rate~\cite{rosakis_intersonic_2002}. To get an idea of how our measured crack velocities compare with $v_R$, we estimate the latter from the expression~\cite{bergmann_ultrasonics_1948,achenbach_wave_1973,meyers_dynamic_1994,quinn2019terminal}

\begin{equation}
v_R=\frac{0.87+1.12\nu}{1+\nu}\sqrt{\frac{G}{\rho}},
\end{equation}

\noindent
where $G$ is the shear modulus, $\nu$ is Poisson's ratio, and $\rho$ is density. (In practice we measured the Young modulus $E$ from the slope of the stress-strain curve at small $\varepsilon$ and $\nu$ from the geometry of the sample under strain. $G$ was then obtained from $G=E/2(1+\nu)$.) In Fig.~\ref{fig1_ss-crack-velo}b we also include the ratio $v_c$/$v_R$ as a function of the Na$_2$O content. One sees that this ratio is high for glasses with low Na content and it decreases significantly with the addition of sodium. This dependence is thus consistent with the view that the fracture process of glasses with low Na concentration is dissipating less energy than it is the case for glasses with high Na content, in agreement with the finding that the former glasses are brittle while the latter are more ductile.

Also of interest is the observation that $v_c$/$v_R$ reaches a maximum at around 9.5\% Na$_2$O (NS10), i.e., the composition at which a significant change in the elastic properties of the sodium silicate glasses is observed~\cite{zhang2022stiffness}. Thus this finding supports the view expressed above that the mechanical response before failure has consequences for the fracture process. We also mention that for the Na-poor compositions, i.e., Na$_2$O\%$~<$~13\%, the estimated $v_c/v_R$ is compatible with the value of $0.5\sim0.6$ that one usually finds in experimental measurements~\cite{quinn_fractography_2007}, while for the Na-rich glasses, $v_c/v_R$ is considerably smaller. The reduction of  $v_c/v_R$ can be related to the fact that the local deviation of the crack front (crack arresting) is more pronounced in these Na-rich glasses, which is also reflected in the change of  fracture surface roughness~\cite{zhang2021roughness}. Finally, we note that all the estimated $v_c/v_R$ are smaller than the experimental upper bound of 0.66 found for inorganic glasses~\cite{schardin_velocity_1959,anthony_crack-branching_1970}.

\subsection{Cavitation}

\begin{figure}[ht]
\centering
\includegraphics[width=0.8\columnwidth]{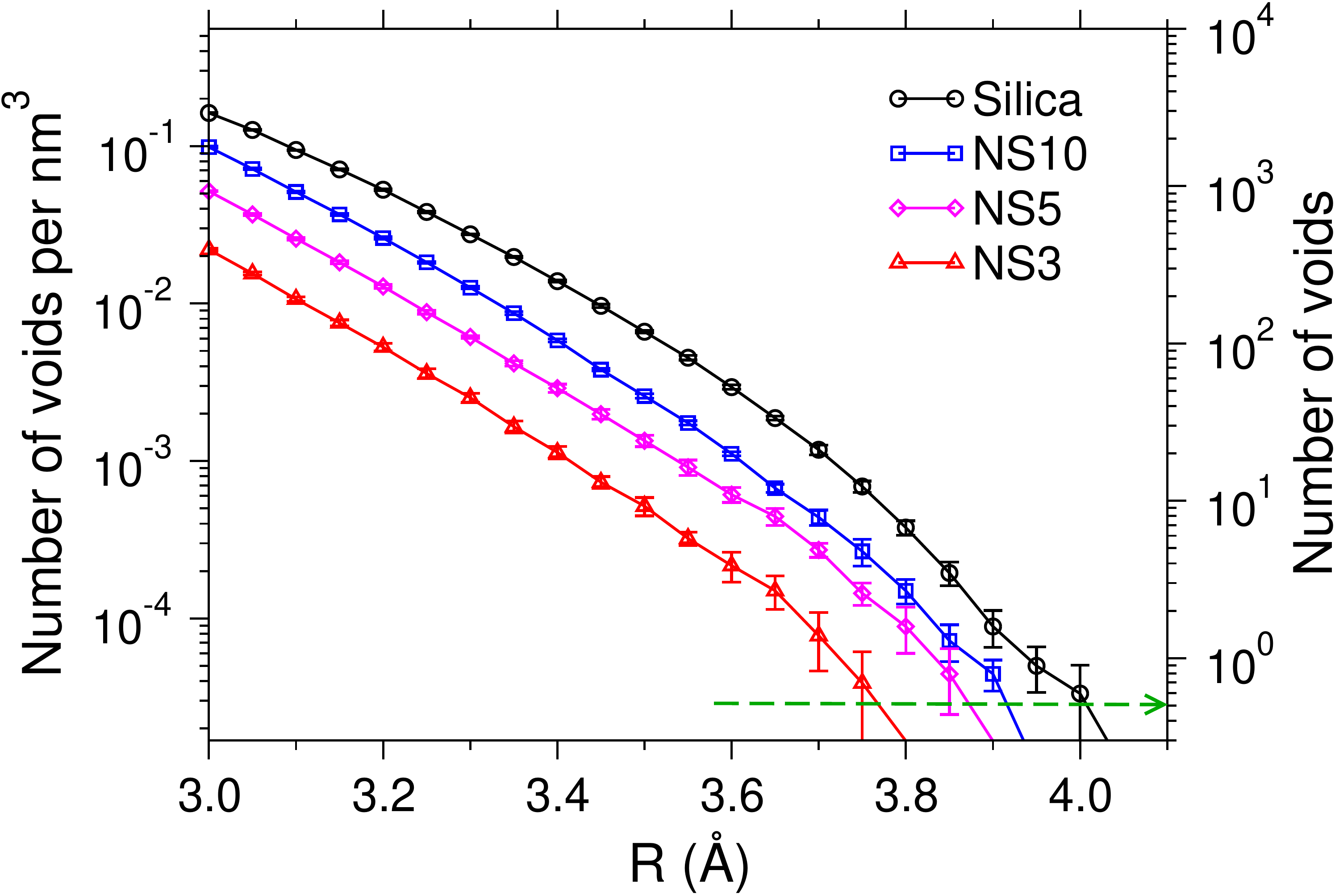}
\caption{Number density (left scale) and number of the voids (right scale) in the unstrained samples versus the radius of the probing sphere, $R$. In the search of voids, the examined samples are slabs of dimensions $20$~nm$\times$30~nm$\times$50~nm. The green dashed line represents $N_{\rm void}$=0.5, i.e., with 50\% probability to find only one void with size $R$ in the unstrained sample. Error bars are standard error of the mean over the six fracture simulations.}
\label{fig2_void-nb-Rc-sdwh}
\end{figure}

To understand the deformation and fracture behavior of the glasses on the microscopic scale, we explored first whether or not glass fracture is accompanied by the formation and growth of cavities. To this aim we need an operational definition of what such a cavity really is since one has to be able to distinguish it from the random voids that are already present in the open network structure of the non-strained glass.
To start, we considered all atoms to be point-like objects, i.e., no volume is assigned to them. Then a 3D cubic grid with a mesh size of 1.25~\AA\ was superposed to the sample and a probing sphere of radius $R$ was moved over all grid points. This mesh size was several times smaller than the probing sphere (see below), thus ensuring that the whole space is properly probed. Those grid points for which no atoms were found inside the probing spherical regions were labeled as empty.  If two neighboring grid points were both classified as empty, the sum of their encompassed probing spherical regions was defined to form one void.  Figure~\ref{fig2_void-nb-Rc-sdwh} shows that for unstrained glasses the void density $\rho_{\rm void}$, i.e., the number of voids per unit volume, decreases basically exponentially with increasing $R$, and that this decrease is independent of the Na concentration. From this data we recognize that the void density decreases by about a factor of 10 if one increases the Na content from 0\% (silica) to 25\% (NS3). This trend can be rationalized to some extent by the fact that the atomic number density $\rho_N$ of the glass increases with the addition of Na$_2$O ($\rho_N = 66.82$ nm$^{-3}$ for silica and $73.83$ nm$^{-3}$ for NS3). However, this modest difference of 10\% is not able to explain
the factor of 10 the density of the void size. Instead this large factor shows that the addition of Na makes that the very open network structure of silica is replaced by a more uniform structure that has significantly smaller holes, in agreement with the observation that the addition of Na leads to a strong decrease of the first sharp diffraction peak in the static structure factor~\cite{horbach_structural_2001}.

From the graph one can conclude that, independent of the Na concentration, voids with a size of $R\approx 4.0$~\AA\ become very rare, i.e., less than one void in the samples we consider, and hence we can use this size as a threshold to distinguish between cavities that are already in the unstrained sample from the ones that are due to the applied stress. We also mention that in a previous simulation study in which the same definition of void was used, it was found that the largest void in unstrained silica glass has a radius of $\approx 3.5$~\AA~\cite{muralidharan_molecular_2005}, i.e., significantly less than the threshold we find here. However, in that work the samples were cubic boxes with a size of $\approx$ 3.7~nm, and our data for silica shows that at $R=3.5$~\AA\  one can expect to find about $10^{-2}$ voids per nm$^3$, i.e., 0.5 voids in a cube of size 3.7~nm, in good agreement with the estimate of Ref.~\cite{muralidharan_molecular_2005}.

\begin{figure}[ht]
\centering
\includegraphics[width=0.8\columnwidth]{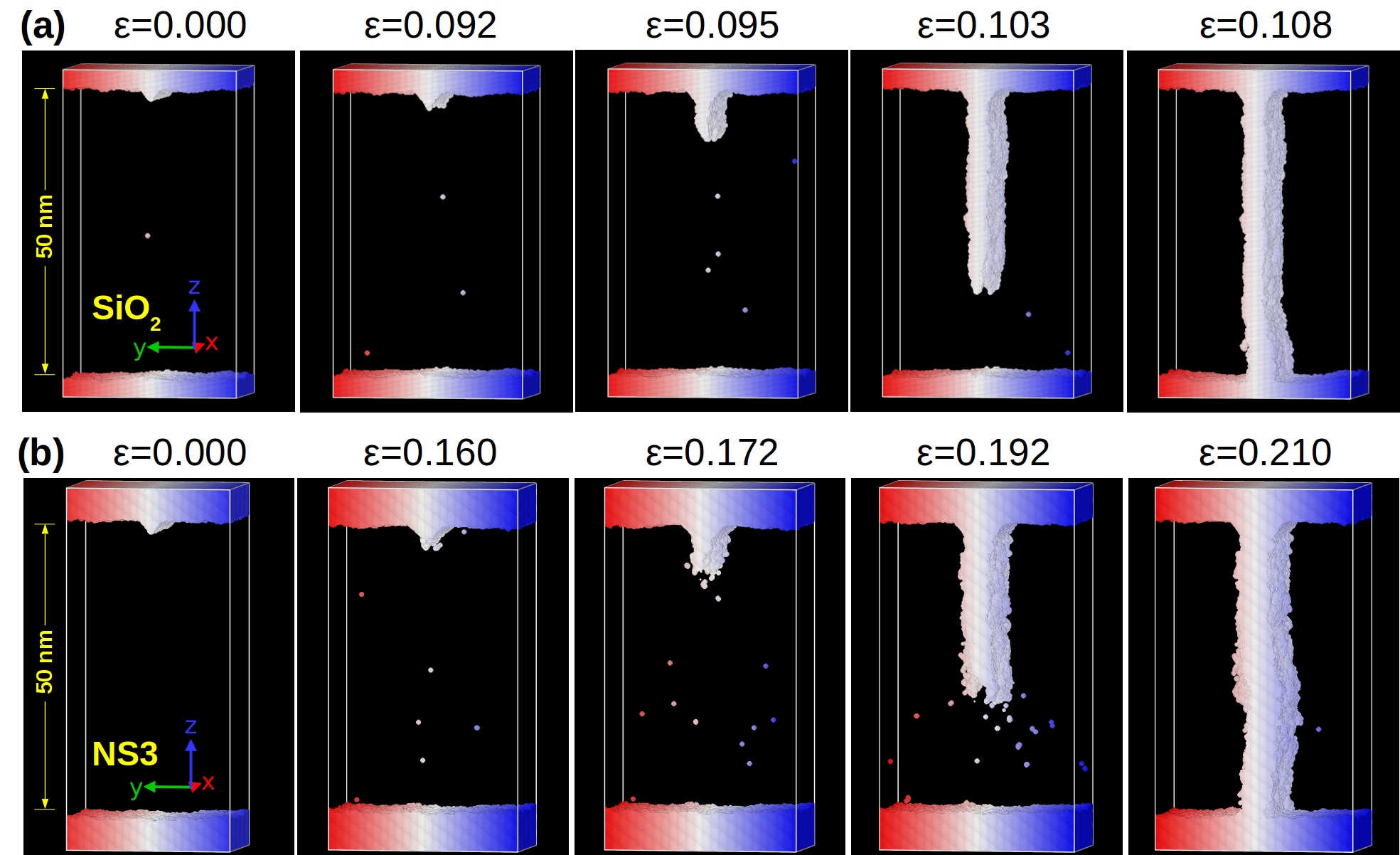}
.
\vspace*{2mm}\\
\includegraphics[width=0.8\columnwidth]{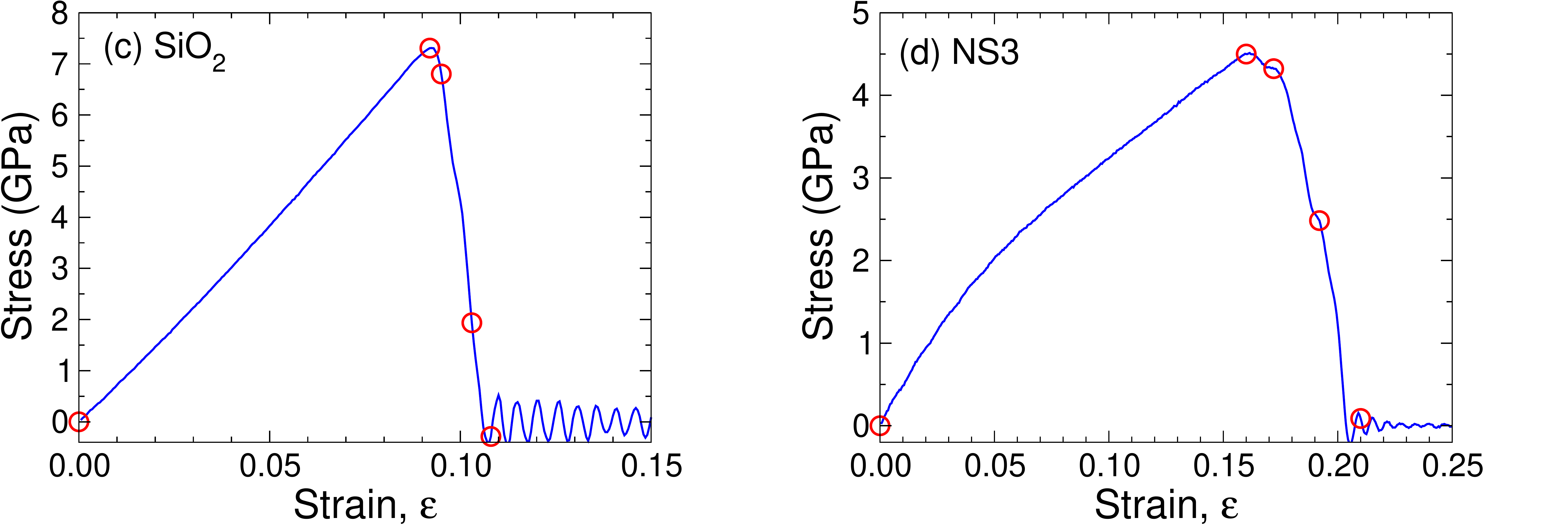}
\includegraphics[width=0.8\columnwidth]{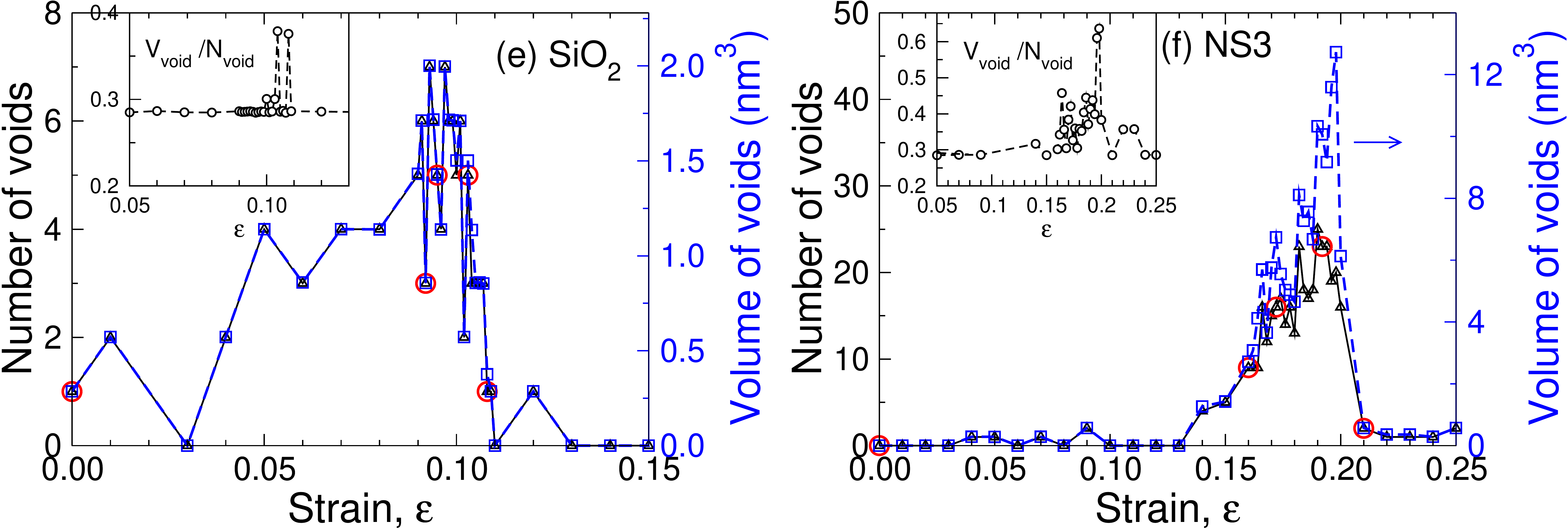}
\caption{(a) and (b): Snapshots showing the empty space during the fracture of silica, (a), and NS3, (b), glasses (sample size $20\times30\times50$~nm$^3$). A notch is introduced on the top surface to initiate the fracture. Color coding is based on the coordinate in the $y-$direction. Movies of the fracture process can be found in Ref.~\cite{movies}. (c) and (d): Stress-strain curves corresponding to the fracture of the two glasses. The red circles indicate the strains at which the snapshots in panels (a) and (b) are shown. (e) and (f): The total number (left scale) and volume of voids (right scale) during fracture of the two glasses. The red circles highlight the specific strain points shown in the snapshots of panels (a) and (b). The insets in panels (e)  and (f) show the ratio between the volume (in nm$^3$) and number of the voids. 
}
\label{fig3_void-snapshots-ss-num}
\end{figure}

Having determined the upper size of the cavities in the unstrained samples, we can now characterize the fracture process of the glasses in terms of voids. To start we present in Figs.~\ref{fig3_void-snapshots-ss-num}(a) and (b) snapshots of the system that show the evolution of the crack front for the silica and Na-rich NS3 glasses, respectively (movies of this process can be found in Ref.~\cite{movies}). Panels (c) and (d) depict the stress-strain curves of the two glasses and  the strains corresponding to the snapshots are marked by circles. The snapshots demonstrate that silica shows typical brittle behavior in that the fracture path is straight and no relevant cavities are found in front of the crack front during the fracture process. (We note that the white dots seen in the snapshots are transient voids which have a size of $R \approx 4$~\AA, i.e., they should be considered as local fluctuations.). To  quantify this last observation, we present in Fig.~\ref{fig3_void-snapshots-ss-num}(e) the number of voids, $N_{\rm void}$, as a function of the strain (left scale) and one recognizes that this number is small and fluctuates strongly. Also included in the graph is the total volume of the voids, $V_{\rm void}$ (right scale), and one sees that this quantity tracks $N_{\rm void}$ perfectly well. The fact, see inset, that the ratio between $V_{\rm void}$ and $N_{\rm void}$ is around 0.286~nm$^3$ during the entire fracture process (except for the two jumps at around 10\% strain due to local structural fluctuations) indicates that all the voids have a size of $R \approx 4.1$~\AA, i.e., they are isolated cavities that have a volume just above the cutoff threshold used to define non-trivial voids, in agreement with the observation from the snapshots. Hence one can conclude that silica glass breaks in a nearly perfect brittle manner by sequential bond breaking at the crack tip, i.e., no cavities are formed ahead of the crack tip. This finding agrees with the experimental results by Guin and Wiederhorn~\cite{guin_fracture_2004} and a recent simulation study using reactive force field~\cite{chowdhury_effects_2019}. 

The scenario is very different for glasses that are rich in Na, such as NS3 shown in panels (b), (d), and (f): Before the crack starts to propagate, i.e., $\varepsilon<0.16$, no stable voids are detected. However, once the fracture process has started, one observes the formation of small cavities close to the crack tip and subsequently these voids grow and merge with the crack front (see panel for $\varepsilon=0.172$). The crack tip also shows micro-branching (see panel for $\varepsilon=0.192$) and this gives rise to a rougher fracture surface. Panel (f) confirms quantitatively the observations from the snapshots: For low strain the number of voids is small and the ratio $V_{\rm void}/N_{\rm void}$, see inset, is basically independent of $\varepsilon$. Once $\varepsilon$ exceeds 0.15, i.e., the crack starts to propagate, the number of voids quickly shoots up and so does the ratio $V_{\rm void}/N_{\rm void}$, indicating that the typical void grow in size.
Hence, one concludes that the fracture of NS3 is accompanied by the formation of nanoscale cavities ahead of the crack front, which subsequently grow and coalescence with the fracture surface. The presence of these cavities is also one of the reasons why the resulting fracture surface has a roughness that is significantly higher than the one found for silica~\cite{zhang2021roughness}.

Although all silica-based glasses are usually considered to be prime examples for brittle materials, the observed differences between the silica and NS3 glasses indicate that the composition does have a crucial role for the fracture behavior on the microscopic scales, in accordance with previous simulations~\cite{wang_intrinsic_2015,pedone_dynamics_2015}. Hence on small length scales silicate glasses with a significantly depolymerized structure can be considered as quasi-brittle materials, i.e., they are not brittle on all length scales~\cite{bauchy_fracture_2016}. (We note here that brittleness will depend to some extent on the quench rate with which the samples has been produced as well as on the strain rate used for the fracture. However, these dependencies are relatively mild and do not affect the present results~\cite{zhang_thesis_2020}.)

\begin{figure}[ht]
    \centering
    \includegraphics[width=0.5\columnwidth]{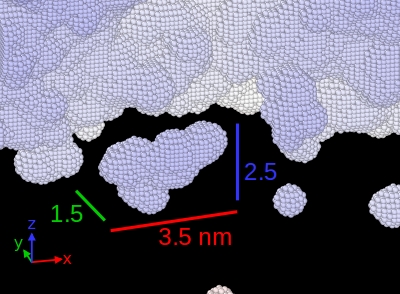}
    \caption{Enlarged view of the largest void found at $\varepsilon=0.196$ for the fracture of a NS3 glass. \\
    }
    \label{fig4_largest-void-ns3}
\end{figure}

A further important question concerns the maximum size of the cavities and their shape. To address these points we have identified the largest void during the fracture of the NS3 glass, see Fig.~\ref{fig4_largest-void-ns3}. It is found that the biggest void has in the $x-$direction (plane parallel to the crack plane, i.e., perpendicular to the tension) an extent of $\approx$3.5 nm, while in the $y-$direction (parallel to the stress direction) the length is only $\approx$1.5~nm. 
This highly anisotropic shape indicates that once a small cavity has formed somewhat ahead of the crack front (since this is where the local stress is highest), it expands in the direction orthogonal to the stress, thus forming a (roughly) lens-like structure. This structure grows and subsequently merges with the approaching fracture front.
Note that once the voids have been absorbed by the passing crack front, it can be expected that they relax elastically in the $y-$direction, at least partially, since the stress is released, i.e., their depth (as seen from the surface) will decrease~\cite{bonamy_nanoscale_2006}.
This recoil might explain why no remnants of voids were detected in an experimental study which compared the two postmortem fracture surfaces of soda-lime-silicate glasses~\cite{guin_fracture_2004}. 

\subsection{Atomic-level properties}
A better understanding of the deformation and fracture behavior of the glass can be obtained by exploring how various atomic-scale properties evolve as a function of strain and how they are related among each other. In this section, we therefore focus on the atomic-scale quantities concerning local compositions, structure, and mechanical properties. More specifically, the local properties considered and their definitions are the following ones: (i) The local number fraction of Na, $f_{\rm Na} = N_{\rm Na}/N_{\rm tot}$, where $N_{\rm Na}$ and $N_{\rm tot}$ are, respectively, the number of Na atoms and the total number of atoms within a given probe volume (defined below). We recall that in NS3 the stoichiometric value of $f_{\rm Na}$ is equal to 0.167; (ii) The local network connectivity as represented by the coordination number of oxygen, $Z_{\rm O}$, i.e., the number of Si in the nearest neighbor shell of a given O; (iii) The local variation of the inter-tetrahedral angle, $\Delta\theta_i(\varepsilon) = \theta_i(\varepsilon)-\theta_i(0)$, where $\theta_i(0)$ is the SiOSi angle of O atom $i$ in the initial configuration (0\% strain), and $\theta_i(\varepsilon)$ is the SiOSi angle of the same O atom in the sample at strain $\varepsilon$.  (Note that it is possible that a SiOSi bridge present at 0\% strain does not exist anymore at strain $\varepsilon$, since one of its Si-O bonds has broken. In this case, we simply assign $\Delta\theta_i(\varepsilon)=0$.);  (iv) The atomic shear strain as first introduced by Shimizu \textit{et al.}~\cite{shimizu_theory_2007} for measuring the local plastic deformation, i.e., a quantity that carries basically the same information as the non-affine squared displacement $D_2^{\rm min}$ introduced by Falk and Langer~\cite{falk_dynamics_1998}; (v) The atomic stress defined via the per-atom stress tensor~\cite{thompson2009general}. (vi) The local temperature as obtained by using the equipartition theorem to map the instantaneous kinetic energy of the atoms in the probing volume to a temperature. 

In order to obtain three dimensional maps of these atom-based quantities we  coarse-grained them by mean of a Gaussian weight function $\phi(r)=A(\omega)\exp[-r^2/\omega^2]$, where $r$ is the distance from the center of the probing sphere and $A(\omega)$ is a normalization factor. 
We have examined the influence of the coarse-graining length-scale $\omega$ on the maps and found that $\omega=8\sqrt{2}$~\AA\ gives stable results that are also consistent with previous simulation studies~\cite{molnar_sodium_2016,wang_nanoductility_2016}. In order
to obtain a visual map that matches the glass matrix, it is useful to get rid of the empty spaces. To this aim we have used the mentioned  coarse-graining procedure to get a map of the mass density and then applied a threshold of 1.5~g/cm$^3$ to define the glass-vacuum interface. This procedure removes close to the surface a layer of about 0.3~nm and the resulting maps match well the shape of the deformed glass sample (see also Ref.~\cite{zhang_thesis_2020}).

\begin{figure}[ht]
\centering
\includegraphics[width=0.98\columnwidth]{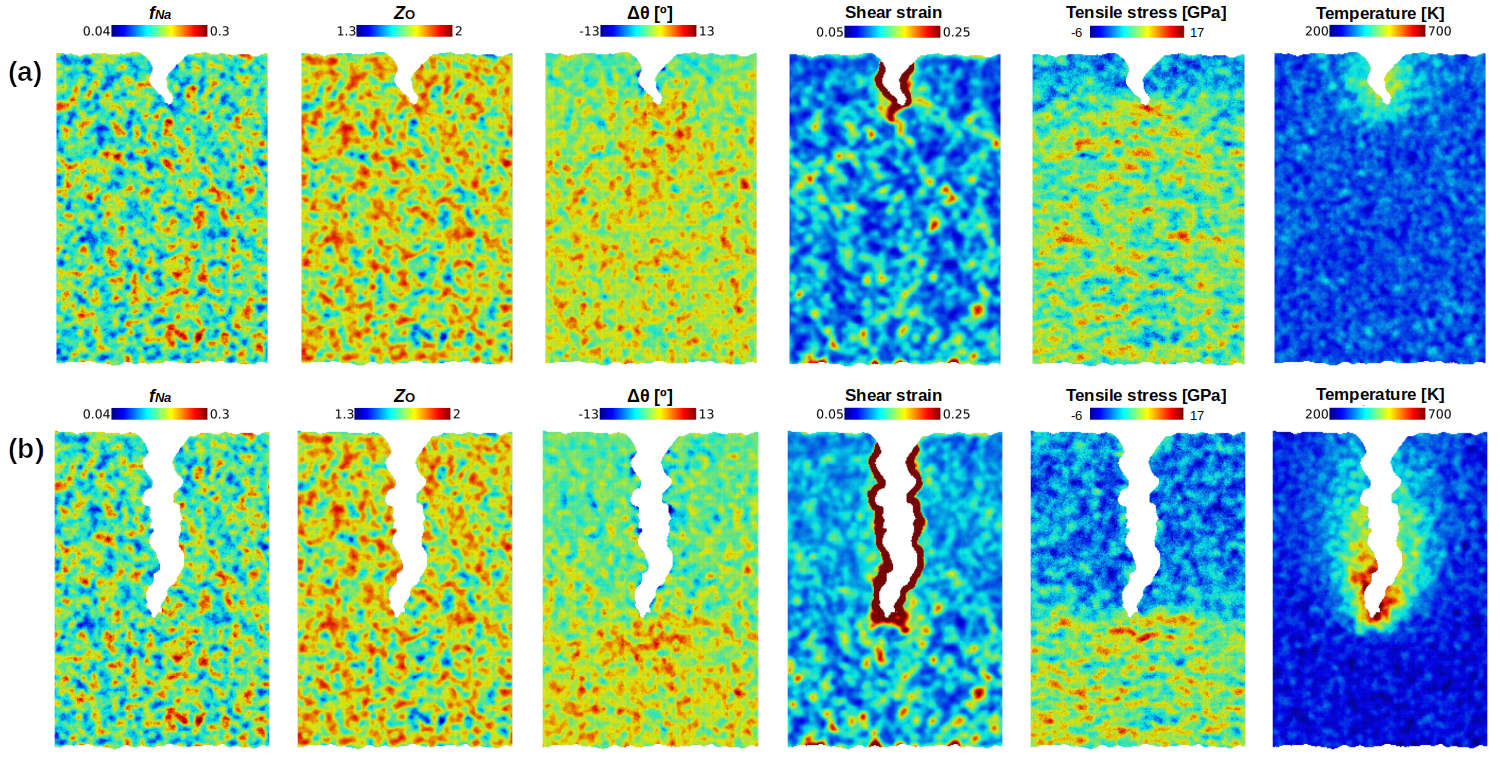}
\caption{Maps of various local properties (see labelling of the panels) at (a) $\varepsilon=0.172$ and (b) $\varepsilon=0.192$ for the NS3 glass. Sample size is $20\times30\times50$~nm$^3$. See the text for the definition of each quantity. These maps are shown for the middle plane of the simulation box in the direction orthogonal to the crack front.
}
\label{fig5_ns3-map-local-properties}
\end{figure}

\begin{figure}[ht]
\centering
\includegraphics[width=0.95\columnwidth]{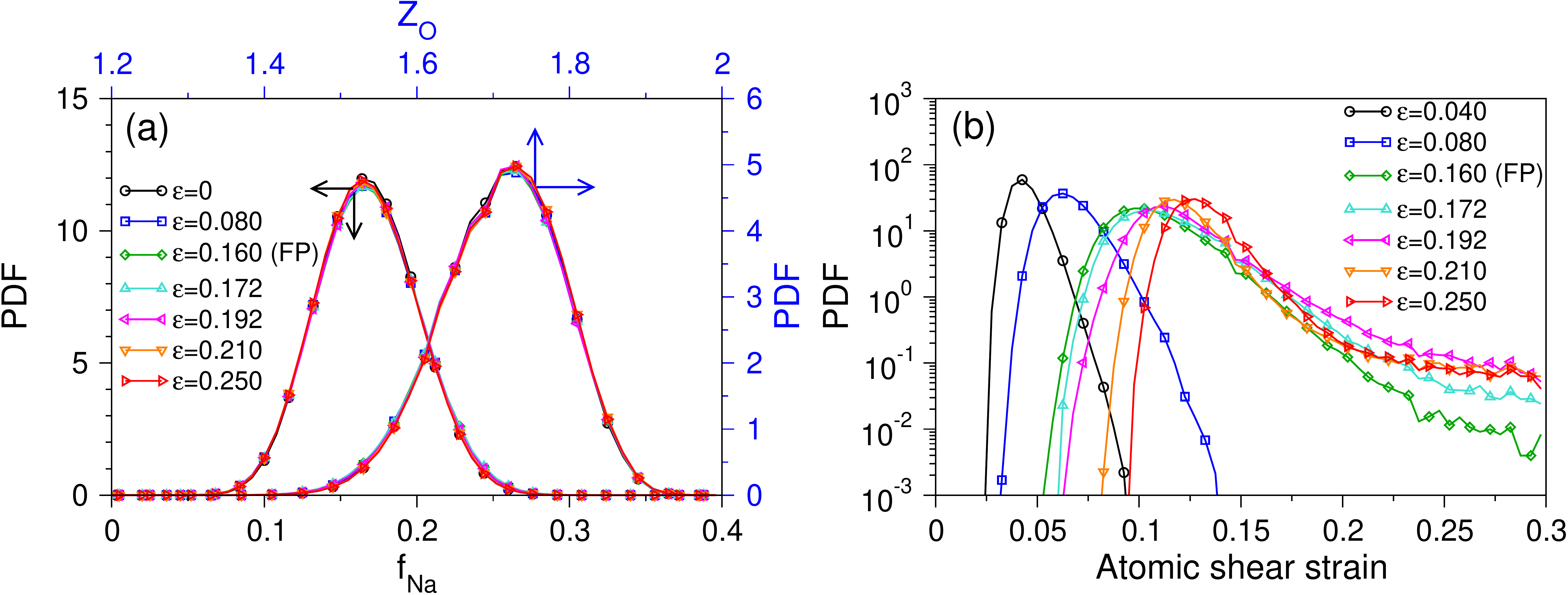}
\includegraphics[width=0.95\columnwidth]{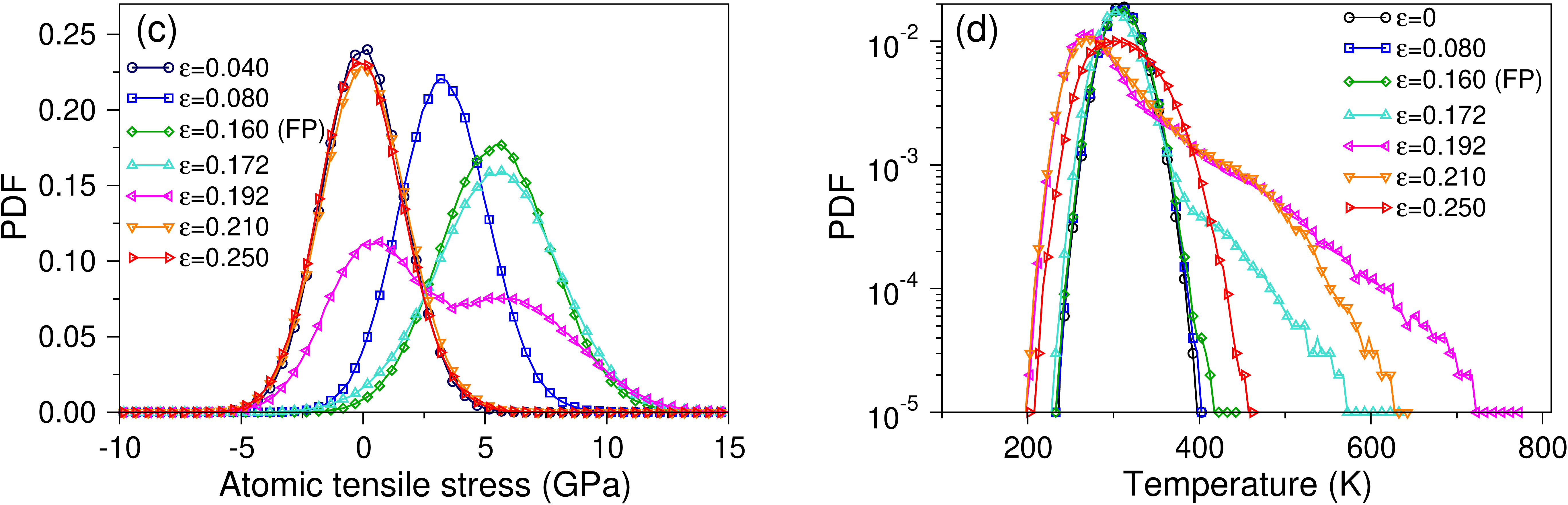}
\caption{Distribution of various local properties in NS3 at various values of the strain. (a) Atomic number fraction of Na (left/bottom axis) and coordination number of O (right/top axis). (b) Atomic shear strain $\eta^{\rm Mises}$. (c) Atomic tensile stress. (d) Kinetic temperature $T$. Note that for this analysis the layers, $\approx2.5$~nm in thickness, near the free surfaces of the sample were not taken into account for calculating the distributions. In panels (b)-(d) ``FP'' denotes the failure point.
}
\label{fig6_ns3-distri-local-prop} 
\end{figure}

Figure~\ref{fig5_ns3-map-local-properties} shows the maps of the various local properties at two representative strains during the fracture of the NS3 glass, i.e., $\varepsilon=0.172$ (the fracture has just started) and $\varepsilon=0.192$ (the fracture front has reached the middle of the sample). (Movies showing the evolution of these local properties during the entire deformation and fracture process are included in Ref.~\cite{movies}.) Firstly, one observes that the local compositional, structural, and mechanical quantities exhibit different degrees of heterogeneities, in that their extent and shape depends strongly on the observable considered. Secondly, a comparison of the snapshots demonstrates that certain properties are correlated to each other. For instance, the local Na concentration $f_{\rm Na}$ and the local coordination number of O, $Z_{\rm O}$, are clearly anti-correlated, and in the vicinity of the crack tip a large change of the SiOSi angle, $\Delta \theta$, seems to be correlated with the large tensile stress. 
Below we will discuss the details of these correlations as a function of the strain. The last column of Fig.~\ref{fig5_ns3-map-local-properties} shows the map of the (kinetic) temperature $T$. One recognizes that close to the crack tip this temperature rises to about 700~K, i.e. several hundred degrees above the average temperature of the sample. This marked local heating, which extends over a distance of 3-7~nm from the fracture surface, is due to the release of the potential energy that was stored in the bonds before fracture and with increasing time this energy will diffuse into the sample. In the context of Fig.~\ref{fig6_ns3-distri-local-prop} we will discuss this process in more detail but already here we mention that this marked local heating can be expected to allow the created surface to relax at least to some extent and thus to become smoother.

To quantify the properties of these local observables, we investigate their distributions as a function of strain. Figure~\ref{fig6_ns3-distri-local-prop}(a) shows that the distributions of local Na concentration and network connectivity are basically independent of the strain, implying that the topology of the local atomic configuration is not changing even if the sample is put under high stress. This observation is in agreement with the results from Ref.~\cite{zhang2022stiffness} where it was shown that up to high strains not only the local topology of the network does not change but that also the alkali atoms are displaced remarkably little ($<3$~\AA), i.e., they will not escape from their surrounding cage.
In contrast to this, Fig.~\ref{fig6_ns3-distri-local-prop}(b) shows that the distribution of the local shear strain $\eta^{\rm Mises}$ depends strongly on the global strain. 
For small $\varepsilon$ the distribution has a narrow peak at small values of $\eta^{\rm Mises}$ (note the logarithmic scale) and it shifts to larger values if $\varepsilon$ is increased, indicating that the sample is increasingly more shear deformed. 
Once the crack starts to propagate, i.e., $\varepsilon \geq 0.16$, one finds that the distribution has a noticeable tail at large $\eta^{\rm Mises}$ which becomes increasingly pronounced with the progression of the crack. This tail arises from the large strains near the fracture surfaces (see the maps in Fig.~\ref{fig5_ns3-map-local-properties}) and thus it can be expected that the weight of this tail is directly proportional to the surface area of the fractured surface. From the graph we also recognize that, once the crack has started to propagate, the main peak of the distribution at around $\eta^{\rm Mises}=0.12$ does not depend anymore strongly on $\varepsilon$. This indicates that there is a maximum amount of local strain that the sample can support before it fractures. Furthermore we notice that in the same range of $\varepsilon$, the distribution is non-zero only if $\eta^{\rm Mises}$ is larger than 0.05, demonstrating that each part of the sample has significant non-affine strain. 

Figure~\ref{fig6_ns3-distri-local-prop}(c) shows the distribution of local tensile stress for different strains. For $\varepsilon=0$ we find that this distribution is symmetric around zero (local) stress and has a full width at half maximum of approximately 5~GPa, indicating that the local stresses are surprisingly high. If the strain is increased, the distribution shifts to the right and at the failure point has a maximum at around 6~GPa. Note that this value is significantly higher than the maximum global stress in the sample, which is, see Fig.~\ref{fig3_void-snapshots-ss-num}(d), around 4.5~GPa. The location of the maximum in the distribution of the local stress does, however, coincide with  the intrinsic failure stress of NS3 as obtained from previous simulation studies using periodic boundary conditions in all three directions, i.e., without a free surface~\cite{zhang_potential_2020,zhang2022stiffness}. In other words, these observations show that the reduction of the failure stress measured here is due to the presence of a surface defect of the sample, i.e., the notch, while the intrinsic failure stress is significantly higher.

The graph also shows that the shift of the distribution to higher stresses with increasing $\varepsilon$ is accompanied with a widening of the distribution, i.e., the local stresses become more heterogeneous, thus reflecting the disorder of the sample on the microscopic scale. 
Once the fracture begins, one observes that the peak position at large stress remains unchanged whereas the peak intensity decreases and the distribution starts to show a weak tail toward small stresses. As the crack advances, this tail is transformed quickly into another peak at small stress and its height increases. Since the maximum of this new peak is located at the same stress as the distribution for $\varepsilon=0$, we conclude that it corresponds to the local stresses in those parts of the sample at which the crack has already passed, i.e., which are no longer under stress, in agreement with Fig.~\ref{fig5_ns3-map-local-properties}(b). For the highest values of the strain the distribution is therefore identical to the one at $\varepsilon=0$, i.e., a globally stress-free state, since the sample has been broken.

The distribution of the local kinetic temperature is shown in panel Fig.~\ref{fig6_ns3-distri-local-prop}(d). For small strains one finds the expected Gaussian distribution peaking at $T=300$~K, i.e., the temperature of the sample. The shape of the distribution changes only once the strain exceeds the fracture point in that it starts to show a pronounced tail to the right. From the temperature map in Fig.~\ref{fig5_ns3-map-local-properties} we recognize that this tail is due to the atoms that are close to the crack front and whose kinetic energy increases rapidly because of the release of the potential energy that was stored in the highly stretched bonds. Note that at these high strains it is not possible to describe the total distribution as a sum of two (or few) Gaussian distributions since one probes here a system that is highly out of equilibrium and therefore the notion of temperature is not well-defined. So the temperature map we show in Fig.~\ref{fig5_ns3-map-local-properties} should be interpreted with some caution. (However, the distribution of the kinetic energy is a well defined quantity.) Only at very large strains, i.e., once the sample is broken, the distribution starts to return back to its original Gaussian shape and it is possible to define again a temperature. The distribution in Fig.~\ref{fig6_ns3-distri-local-prop}(d) also shows that for strains that are just above the failure strain, e.g. $\varepsilon=0.192$, the location of the main peak 
is at around 270~K, i.e., significantly below the temperature of the unperturbed sample, in agreement with the fact that in the temperature map in Fig.~\ref{fig5_ns3-map-local-properties} the regions far away from the crack front do have a low $T$. This result might look odd since for distances far away from the crack front one expects that the local temperature is the mean temperature of the sample. The explanation for this somewhat unexpected behavior is the fact that in the simulation we use a thermostat which makes that the total kinetic energy of the particles is fluctuating around the fixed target value. Thus the fact that a small portion of the sample becomes very hot, will automatically make that the rest of the sample will have an average kinetic energy, and thus temperature, that is below the one of the target value, thus explaining why the regions of the sample far away from the crack front are colder than expected. We emphasize, however, that this finite size effect does not influence in a significant manner the interpretation of the temperature map shown in Fig.~\ref{fig5_ns3-map-local-properties}.

\begin{figure}[ht]
\centering
\includegraphics[width=0.95\columnwidth]{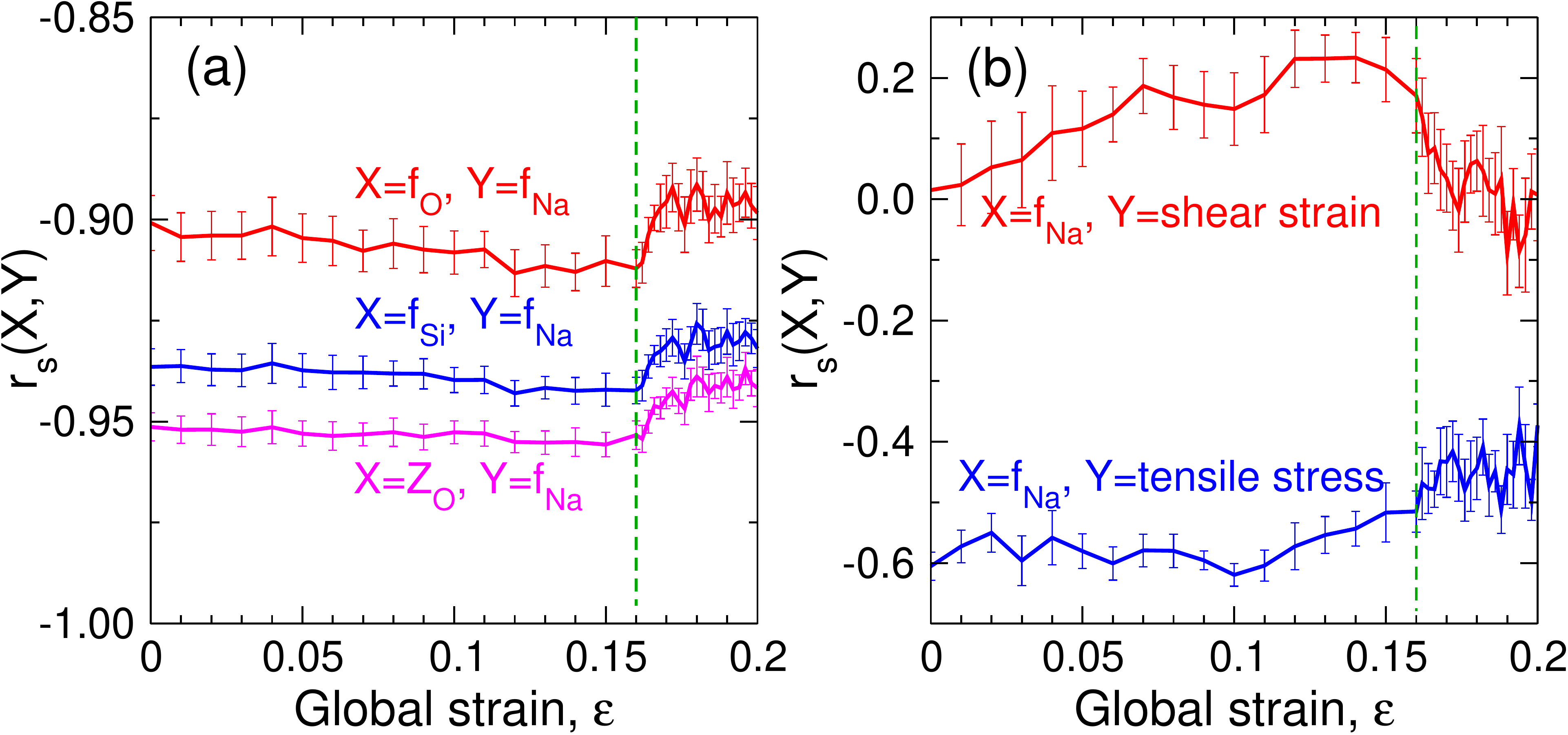}
\includegraphics[width=0.95\columnwidth]{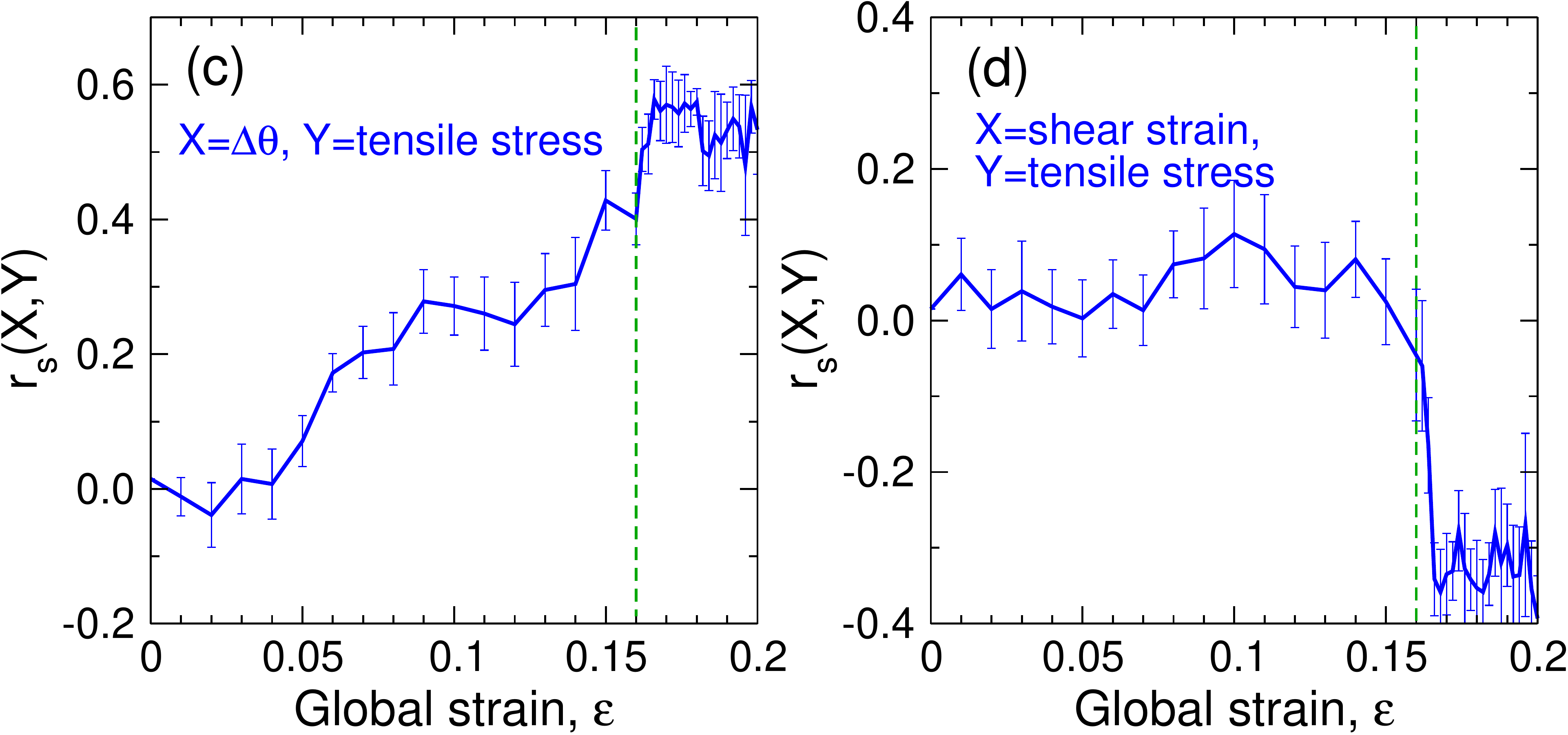}
\caption{Spearman's correlation coefficient, $r_s$, between various local properties of the NS3 glass, see labeling of curves for details.
The vertical dashed lines indicate the failure point (i.e., at maximum stress). Error bars are standard error of the mean of 10 planes that are evenly spaced orthogonal to the direction of the crack front, i.e., orthogonal to the $y$-axis. 
}
\label{fig7_ns3-corr-sdwh-local-properties}
\end{figure}

Further insight into the mechanical properties of the glass can be obtained by quantifying the strain dependence of the correlations between the local quantities. Since the propagation of the crack depends strongly on the local structure of the material structure close to the tip, we will focus on this region for characterizing the correlations between the various local quantities. In practice we considered a plane of the sample orthogonal to the $x-$axis, i.e., orthogonal to the crack front,  and probed a circular region around the crack tip having a radius of 5.0 nm, i.e., a size that is comparable to the extent of the process zone as estimated in previous dynamic fracture experiments of silicate glasses~\cite{pallares2012fractoluminescence}. (We note that the size of this zone, as determined from the temperature increase close to the crack tip, depends only very weakly on glass composition and crack velocity, i.e., the chosen size is not a critical parameter for our analysis~\cite{zhang_thesis_2020}.) 
 
To quantify the correlation between various local properties we use the Spearman rank correlation coefficient, $r_s(X,Y)$, which is a measure for the statistical dependence between two variables $X$ and $Y$~\cite{spearman_proof_1901}. By construction $r_s$ is in the range [-1, 1] and $r_s>0$ means positive correlation, i.e., $Y$ tends to increase when $X$ increases, and $r_s<0$ indicates an anti-correlation. The magnitude of $r_s$ gives the degree of correlation, with $r_s=0$ indicating no correlation between the two variables.

To start, we show in Fig.~\ref{fig7_ns3-corr-sdwh-local-properties}(a) the correlation between local elemental concentrations. One observes that Si and Na concentrations are strongly anti-correlated in that $r_s({\rm Si,Na})\approx -0.94$. A strong anti-correlation is also found between O and Na with $r_s({\rm O,Na})\approx -0.90$, in agreement with previous studies on soda-silicate glasses~\cite{molnar_sodium_2016}. 
From panel (a) one also recognizes that $f_{\rm Na}$ and $Z_{\rm O}$ are strongly anti-correlated, i.e., Na-rich regions have a smaller network connectivity in agreement with the expectation that the presence of an alkali atom leads to the breaking up of the network. A slight reduction of the correlations can be seen during the fracture process, which is likely due to the enhanced local disorder near the crack tip, arising from the local stress-and heat-driven rearrangement of atomic positions.
Overall, we note that the curves in panel (a) are basically independent of the strain (both before and during fracture), which indicates that the local compositions and network connectivity does not change in a significant manner even if the glass is under strong deformation.

Figure~\ref{fig7_ns3-corr-sdwh-local-properties}(b) shows the correlations between local Na concentration and mechanical properties. Overall, we note that $f_{\rm Na}$ is positively correlated with local shear strain, and anti-correlated with local tensile stress. The former result can be rationalized by recalling that Na-rich regions are more flexible due to the high mobility of Na and the reduced connectivity of the network. Due to this flexibility these regions can deform relatively easily and hence reduce stress, in agreement with the observed anticorrelation.
We point out, however, that the absolute value of $r_s$ that quantifies the correlation between $f_{\rm Na}$ and $\eta^{\rm Mise}$ is quite small. This indicates that the local plastic deformation cannot be solely attributed to the enrichment of Na since it depends also on other factors such as, e.g., the rotational motion of a tetrahedron. In contrast to this, the correlation between $f_{\rm Na}$ and local tensile stress has a substantially stronger correlation, $r_s \approx -0.6$ at the early stage of the deformation and $r_s \approx -0.5$ during fracture, in agreement with the maps shown in Fig.~\ref{fig5_ns3-map-local-properties}. Thus this indicates that the local stress level is mainly related to the flexibility of the structure which in turn is strongly influenced by the presence of Na.

The correlation between the local change of the SiOSi angle and the local tensile stress is shown in Fig.~\ref{fig7_ns3-corr-sdwh-local-properties}(c). We observes that the correlation between these two quantities gradually builds up (basically linearly), with $r_s$ increasing from close to zero at small strains to around 0.55 for $\varepsilon$ beyond the failure strain. 
The strain dependence of this correlation can be understood by considering the coexistence of multiple deformation modes during tension: As it was shown in Ref.~\cite{taraskin_connection_1997} by probing the vibrational density of states, the softest modes are the rotations of the tetrahedra and the change in the SiOSi angle. Hence it can be expected that at low and intermediate strains these two modes become active and therefore one finds an increasing correlation between stress and the change in the SiOSi angle. In contrast to this the change in the Si-O bond length with increasing strain is small, see Ref.~\cite{zhang2022stiffness}, and hence one does not expect a significant correlation of this length with the value of the stress. Thus, the SiOSi angle is likely the simplest local structural quantity that is sensitive to the local stress.

Finally we present in Fig.~\ref{fig7_ns3-corr-sdwh-local-properties}(d) the correlation between local shear strain and tensile stress and one finds that at small and intermediate strains there is basically no correlation between these two quantities. 
This result can be rationalized by recalling that on the microscopic scale the mechanical properties of a glass are quite heterogeneous~\cite{zhang2022stiffness,wang_intrinsic_2015,molnar_sodium_2016,wang_nanoductility_2016}, i.e., there are soft spots as well as regions that are relatively stiff, see Fig.~\ref{fig5_ns3-map-local-properties}. One can expect that the application of an external stress will make that the soft regions
will respond easily to an applied stress, thus releasing it by rearranging the particles, i.e., the local strain is large. In contrast to this, stiff regions will not change their geometries, i.e., the strain is small. Since, however, at the interface of these two type of regions the local stresses need to be, more or less, balanced for the sake of mechanical stability, at the end the residual stress in the soft regions are not significantly smaller than the ones in the stiff regions, making that there is no substantial correlation between stress and strain, in agreement with the data in panel (d). This argument is no longer valid once the sample has reached its failure point, since in this case the presence of a propagating crack makes that there is no longer a local stress balance. As a result, one finds that there is a notable anti-correlation between local stress and shear strain with a correlation coefficient of around -0.35 during crack propagation.

\section{Summary and conclusions} \label{summary}
To summarize, we have investigated the dynamic fracture of sodium silicate glasses under tension by using large scale molecular dynamics simulations. This study highlights the strong influence of the composition on the fracture behavior of glasses on the microscopic scales. The stress-strain curves demonstrate that while silica glass shows a nearly perfect brittle fracture behavior, the Na-rich glasses do exhibit a certain degree of ductility during the deformation and fracture process. A detailed analysis of void formation during the fracture process reveals that no growth and coalescence of voids are observed for silica glass, while
the fracture of the Na-rich NS3 glass is accompanied by the nucleation of nanoscale voids, which subsequently grow and merge ahead of the crack tip. We  have found that, for the NS3 glass, the linear dimension of the voids depend on the direction and that their largest extent is on the order of several  nanometers. Further insight into the fracture  of the glasses  has been obtained by exploring the spatial and temporal evolution of various atomic-level properties and the correlations between them. By probing the evolution of local kinetic temperature during fracture, we find that close to the crack tip the kinetic temperature rises several hundred degrees above the average temperature of the sample, an increase that is observed in a region of size $\approx 10$~nm around the fracture tip.

As a final remark we note that the fracture of glass is obviously a subject of high complexity since many factors, such as the length scale under investigation, the strain rate, and the chemical composition/microstructure, play an important role. This complexity might make it impossible to come up with universal conclusions regarding the fracture behavior of silicate glasses. Thus, although the present work provides a microscopic perspective on the dynamic fracture behavior of sodo-silicate glasses, it remains to be explored how other glass-formers used in practical applications, such as boro-silicates or alumino-silicates, will fracture on the microscopic scale since the latter systems have several kinds of network-formers which can be expected to influence the fracture dynamics. Further studies on more complex chemical compositions and mechano-chemical conditions will also be valuable to obtain a more comprehensive understanding of the failure of glasses, and for the future design of glassy materials with improved mechanical performance.

\section{Acknowledgments}
Z.Z. acknowledges financial support by China Scholarship Council (Grant No. 201606050112). W.K. is member of the Institut Universitaire de France. This work was granted access to the HPC resources of CINES under the allocations (Grant Nos. A0050907572 and A0070907572) attributed by GENCI (Grand Equipement National de Calcul Intensif).

Z.Z. is currently at College of Materials Science and Engineering, Xi'an Jiaotong University.

\normalem  

\onecolumngrid

\end{document}